\documentclass[10pt]{iopart}
\usepackage{amssymb}
\usepackage[dvips]{graphicx}
\begin{document}
\title[Relativistic and Quantum Properties of Spatial Rotations]{Relativistic and Quantum Properties of Spatial Rotations} 
\author{ A.V.Novikov-Borodin\dag\ }
\address{\dag\ Institute for nuclear researches of the Russian Academy of Sciences, \\ 117312 Moscow, 60-th Anniv. pr. 7a, Russia.}
\ead{novikov@inr.ru}
\begin{footnotesize} \hspace {100mm} \today \end {footnotesize} 
\begin{abstract} 

The principles of the physical description of non-inertial frames of reference are analyzed. The systems of physical reality description (PhRD) are introduced on base of generalization of the relativistic principle in special and general theories of relativity. Physical objects and their interactions from different systems of PhRD are examined. A lot of conformities between considered objects and their interactions with known physical objects and fundamental interactions both in micro- and macro-scales are found. 

\end{abstract}
\pacs{\\ 03.65. Bz - quantum mechanics, foundations; 03.30. + p - special relativity; \\ 11.30. Cp - Lorentz and Poincar \'e invariance in particles and fields; \\ 12.90. + b - Miscellaneous theoretical ideas and models of elementary particles; \\ 04.20. Cv - Fundamental problems and general formalism in general relativity.} 
\section{\bf Principles of the physical description of non-inertial frames of reference.}
\label{sec:PhRD}

It is known from the special theory of relativity, that laws of physics are invariant concerning inertial frames, i.e. moving rectilinearly and uniformly in ``empty'' space. Such formulation means a relativity of movement, but does not give an answer to a question: concerning what movement in ``empty'' space can be determined? How to choose the inertial system? 

It is possible to find some answers to these questions from the principle of relativity of Poincar\'e \cite{Poin04} generalized by A.Logunov \cite{Log89} as follows: ``For any physical frame of references (inertial or non-inertial), it is always possible to select the system of other frames, in which all physical processes are going similar to the initial frame, so we do not have and it is impossible to have any experimental possibilities to distinguish \ldots what, exactly, frame from this infinite system we are in''. 

This approach leads to another logical difficulty: which one from two non-inertial to each other observers may be considered as an inertial one? The principle of relativity doesn't give a criteria to choose. 

It seems that in general theory of relativity the problem is implicitly solved more particularly. Actually, the principle of equivalence of inertial and gravitational masses connects the space-time topometry with existing physical objects (gravitating bodies, particles, fields). The movement over geodesic lines is considered instead of inertial motion, so inertial motion is defined by the topology, by the distribution of matter in the Universe. The World is such what it is. At first sight it seems that this is an answer, but is it really? 

The first of all, the principle of relativity of GTR is valid both for any local region (in micro-scale) and far from the gravitating bodies (in macro-scale). It is one of the basic GTR principles. At least for these regions the problem is not solved. 

Another and more principal point may be formulated as follows. Let's consider again two non-inertial to each other observers from the point of view of Logunov's principle. The space-time properties will be quite different for them. It will be shown in the following analysis that even observable regions may not coincide with each other for non-inertial observers. Does the same ``physical world'', the same ``physical objects'' every observer see? If not, every observer has chance to build his own space-time topometry based on observable by him physical objects (gravitating bodies, particles, fields). So, we still don't have a criteria to choose. 

It is incredible, but it exists a possibility to generalize the Logunov's and GTR approaches to relativity. On base of this generalization some decisions of the formulated problem may be gotten. 

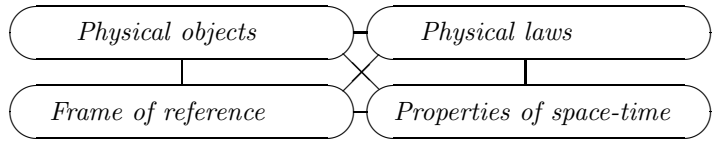
\begin{figure} [h]
	 \begin{picture} (450,75)
	 \put (100,30) {\oval (130,20)} \put (50,27) {\it Frame of reference}
	 \put (100,60) {\oval (130,20)} \put (60,57) {\it Physical objects}
	 \put (235,30) {\oval (130,20)} \put (180,27) {\it Properties of space-time}
	 \put (235,60) {\oval (130,20)} \put (190,57) {\it Physical laws}
	 \put (100,40) {\line (0,0) {10}} 	 \put (230,40) {\line (0,0) {10}}
	 \put (165,30) {\line (1,0) {5}} 	 \put (165,60) {\line (1,0) {5}}
	 \put (161,38) {\line (1,1) {13.5}} 	 \put (161,52) {\line (1,-1) {13.5}}
	 \end{picture}
	 \caption{System of the physical reality description.}
	 \label{fig:PhRD}
\end{figure}

We shall guess, that the physical reality may be considered by any observer only within the framework of some chosen system of the physical reality description (system of PhRD) (see. Fig.\ref{fig:PhRD}). The choice of this or that system of PhRD is defined in connection with the physical objects existing in it. The observer by himself is a part of his own system of PhRD. 

The suggested approach has far-reaching consequences. For example, there is probable an existence together with our system of PhRD a lot of other systems with their ``own'' physical objects, physical laws and properties of space-time. Consideration of various systems of PhRD essentially differs from studying of inertial or non-inertial movement of physical objects inside some fixed system of PhRD. For example, with the description of non-inertial movement the general theory of relativity can be quite consulted. 

Our subject is an investigation of physical objects from various systems of PhRD, their properties and an opportunity of their observation from other systems. We need to note at once , that if such objects really exist, their properties should not ``considerably'' contradict to authentically known physical laws and need to have correspondences with some known physical objects in system of the observer. We shall name this principle a principle of compatibility for systems of PhRD. 

On basis of the suggested approach, we also shall consider systems of PhRD for spatial rotation which further we shall name simply systems of space (or spatial) rotation. 

\section{\bf Structure of spatial rotation objects.}
\label{sec:Struct}

The metrics in frame $K'$ uniformly rotating in space around some spatial axis (we shall consider, without loosing generality, an axis $0z$) with frequency of rotation $\omega_z =\Omega_z  c$, where $c$ is speed of light in system of the observer $K$, in cylindrical coordinates $ (\rho, \phi, z) $ can be determined (see, for example, L.D.Landau, L.M.Lifschitz \cite {LL88}) as:  

\begin{equation}
\left( ds' \right)^{2} = \left( 1- \Omega_z^2 \rho^2 \right) d\tau^2 \pm 2\Omega_z\rho^{2}d\phi d\tau -\rho^{2} d\phi^{2} - d\rho^{2} - dz^{2}. 
\label {ASRint}
\end{equation}  

It is considered, that this expression can be used up to some distance (we shall name it further the horizon of events) $\rho_h = c/\omega_z=1/\Omega_z $ from an axis of rotation determined in system of the observer, otherwise the factor $g_{00}$ (the factor at $d\tau^2 $) of metric tensor will be negative, that is considered inadmissible. 

In respect to the approach suggested in the previous section, the system of PhRD can be created for any combination of spatial rotations. For simplicity we shall consider a case at which axes of rotation from a combination of rotations at least have one common point which we shall name the common point or the center of rotation. In this case in system of the observer $K$ and in system of rotation $K'$ it is possible to write down possible transformations of spatial coordinates $X = (x, y, z) = (x^1, x^2, x^3) = (x^{\alpha})$, $\alpha=1,2,3 $  with help of expressions: 
\begin{equation} 
\fl \quad X'= XR,\: R^{MSR}= R_1^{ASR}\cdots R_n^{ASR},\:R^{SSR}=R_1^{MSR}+\cdots+R_n^{MSR}. 
\label{coord}
\end{equation}
Here $R$ is a matrix of rotations. The matrix of axial rotation $R^{ASR}$ (axis space 
rotation) is a matrix of transformation of coordinates for rotation around some spatial axis\footnote{For example, the matrix of spatial rotation around $0z$-axis in the cartesian coordinates looks like: 
$ R^{ASR}_{z}(\tau) = \left( \begin{array} {ccc} 
 \cos ( \Omega_{z} \tau) & \mp\sin ( \Omega_{z} \tau) & 0 \\ 
  \pm \sin ( \Omega_{z} \tau ) & \cos ( \Omega_{z} \tau) & 0 \\ 
 0 & 0 & 1 \end{array} \right).$}. 
The matrix $R^{MSR}$ (multiple SR) is a matrix of a combination of axial spatial rotations with multiplied matrixes of transformation and the matrix $R^{SSR}$ (sum SR) is a matrix with summable matrixes of MSR transformations. 

The metrics in system of rotation $K'$ from the point of view of system of the observer $K$ we shall introduce by means of an interval: 

\begin{equation}
\fl (ds')^2 = (d\tau)^2 - \| dX'  \|^2, \: dX' = X dR + dX R, \: \| dX' \|^2= dX' dX'^T,  
\label{metrics}
\end{equation} 

where $X^T$ means transposing. In case of a combination of rotations with the common point $ \left( R = R(\tau) \right) $, we shall receive an expression: 
\begin{equation}
\fl (ds')^2 = \left( 1- X \frac{\partial R}{\partial \tau} \frac{\partial R^T}{\partial \tau} X^T  \right) d\tau^2 - dX R R^T dX^T - 2 dX R \frac{\partial R^T}{\partial \tau} X^T d\tau. 
\label{SRmetr}
\end{equation}

It is easy to check up, that for axial rotation with $R (\tau) =R^{ASR}_z(\Omega_z, \tau) $ the  expression for metrics obtained with (\ref{metrics},\ref{SRmetr}) completely coincides with (\ref{ASRint}). 

Expression (\ref{SRmetr}) determines elements of metric tensor in system of the observer for spatial rotations. In case of MSR and SSR the elements of metric tensor depend on time, we shall examine their values as average on time of observation. If time of observation is much greater the period of any considered rotation from a combination of rotations (that, certainly, is more typical for micro-scale analysis), for intervals $MSR$ with a matrix of transformation $R^{MSR}(\tau) = R_x^{ASR}(\Omega_x) R_y^{ASR}(\Omega_y) R_z^{ASR}(\Omega_z)$ and $SSR$ with $R^{SSR}(\tau) = R_x^{ASR}(\Omega_x) + R_y^{ASR}(\Omega_y) + R_z^{ASR}(\Omega_z)$ we shall receive: 

\begin{eqnarray}
\fl \langle \left( ds'_{MSR} \right) ^2 \rangle_T = \bigg\{ 1 - \bigg[ \rho^2 \Omega_z^2 + \left( \frac{1} {2} \rho^2 + z^2 \right) \Omega_y^2 +\left( \frac {1} {2} z^2 + \frac{3}{4} \rho^2 \right) \Omega_x^2\bigg] \bigg\} d\tau^2 +\nonumber \\
\fl \qquad + 2 \rho^2 \Omega _ {z} d\phi d\tau - \rho^2 d\phi^2 - d\rho^2 - dz^2, 
\label{MSRint}
\end{eqnarray} 
\begin{eqnarray}
\fl \langle \left( ds'_{SSR}\right)^{2}\rangle_{T} = \bigg\{ 1-\bigg[ \left( x^2+y^2 \right) \Omega_{z}^2 + \left( y^2+z^2 \right) \Omega_{x}^2 + \left( x^2+z^2 \right) \Omega_{y}^2 \bigg] \bigg\} d\tau^2 -\nonumber \\ 
 \fl \qquad -2 \left[ \left( ydx-xdy \right) \Omega_{z}- \left( ydz+zdy \right) \Omega_{x} + \left( zdx-xdz \right) \Omega_{y} \right] d\tau - 3 \left( dx^2+dy^2+dz^2 \right) \nonumber. 
\end{eqnarray}

The expression for MSR (\ref{MSRint}) is submitted in cylindrical coordinates to emphasize the presence of some axis allocated in space (in the given example it is an axis $0z$). SSRs generally have no such axis allocated in space. 

The approach proposed in section \ref{sec:PhRD}, allows to expand treatment of the metrics in systems of rotation. Physical objects of rotation are not obliged to be at all only inside the horizon of events. This requirement is necessary for physical objects from system of PhRD of the observer. Physical objects of rotation from point of sight of other system of PhRD we shall subdivide on observable, taking place inside horizon of events in system of the observer ($g_{00}> 0 $) and non-observable ones, taking place outside ($g_{00} < 0 $). 

\begin{figure} [h]
	 \begin{center}
		 \includegraphics*[angle=270, width=60mm] {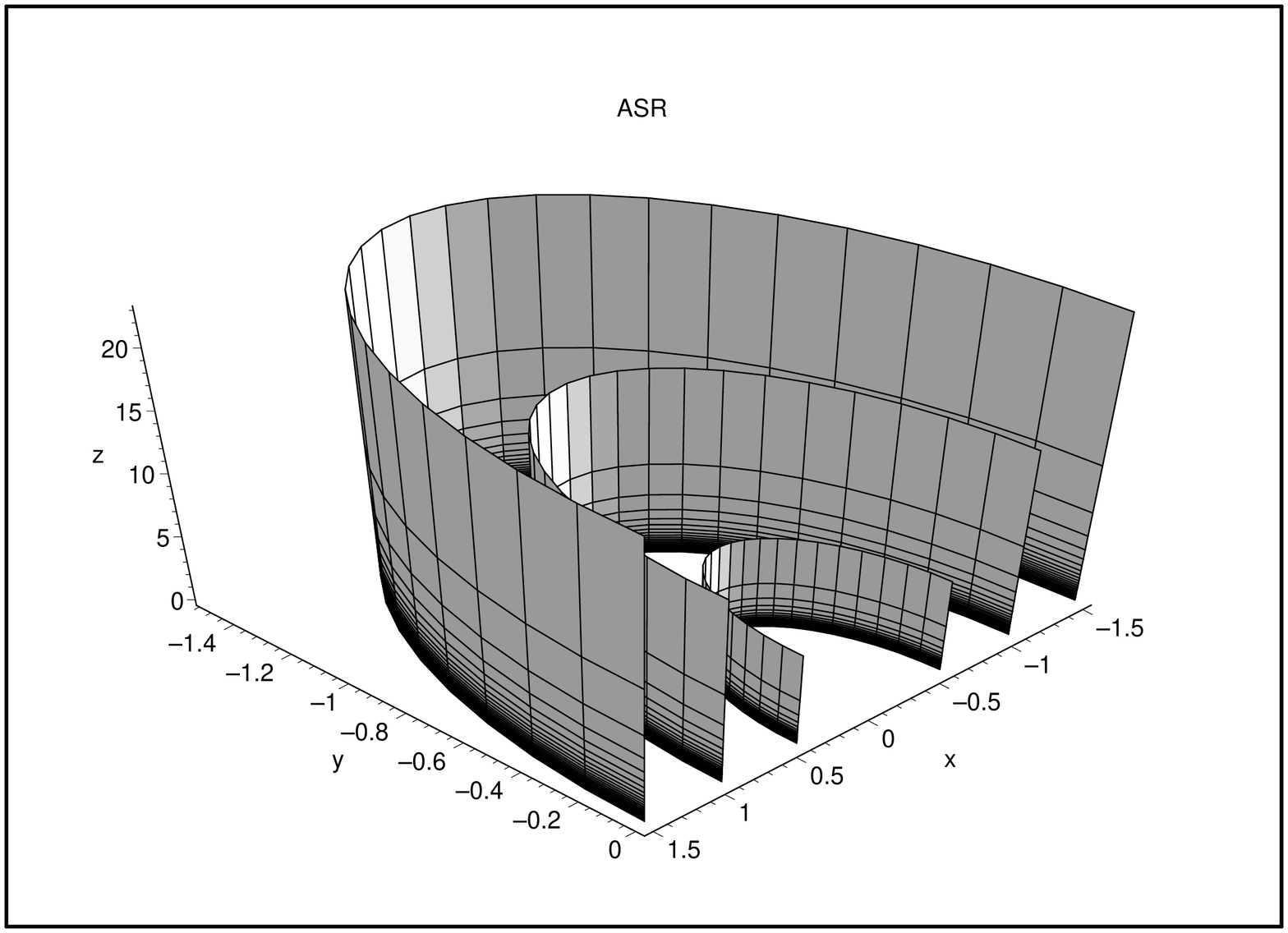}
		 \includegraphics*[angle=270, width=60mm] {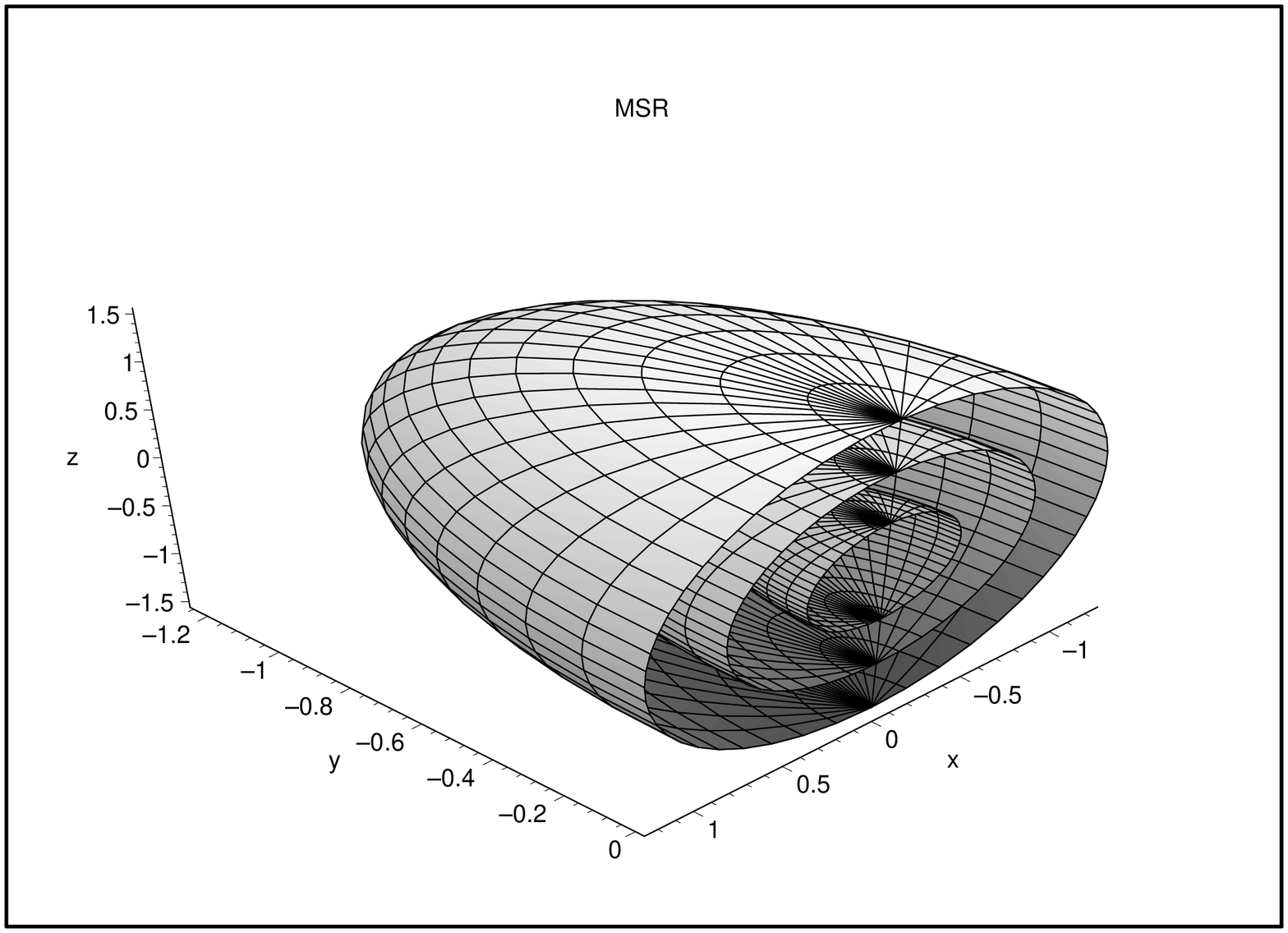}
	 \caption{The ASR and MSR stable regions.}
	 \label{fig:SRobj}
	 \end{center}
\end{figure}

The equation $\langle g_{00} \rangle_T = 0 $ determines a conditional spatial border between observable and non-observable objects of rotation in system of PhRD where this equation is determined. In case of axial rotation it will be a set of cylindroids around of an axis of rotation (at various frequencies of rotation), and in case of MSR and SSR  a set of ellipsoids (see Fig.\ref{fig:SRobj}). 

From the point of view of the external observer the proper time in system of rotation tends to infinity at approach to horizon of events, since at $g_{00} \rightarrow 0: $ $ \Delta\tau =-\oint (g_{0\alpha}/g_{00}) dx^{\alpha} [ \stackrel {ASR} {=} \oint \Omega \rho^2 / (1-\Omega^2\rho^2) d\phi ] \rightarrow \infty $ \cite {LL88}. So, observable objects of rotation seem ``captured'' inside the area limited by horizon of events. As the signature of the metrics in internal area coincides with the signature of the metrics of the observer, physical objects of rotation in this area can ``be identified'' in system of the observer and their properties can be investigated by ``visible'', ``strong'' interactions with others, already known physical objects from system of the observer. 

It is hardly possible to identify the non-observable objects of rotation in system of the observer in connection with change of existential perception. Some opportunities of the description of such objects are suggested in the following section. 

\section {\bf The Description of objects of rotation}
\label {sec:Descript}

Let's the function $ \psi'(X') $ corresponds to some object of rotation in system of PhRD $K ' $. We shall try to find function $ \psi (X, \tau) $, determined in system of the observer $K $ and corresponding to $ \psi ' $: $ \psi (X, \tau) \rightleftharpoons \psi ' (X ') $. Attempts to find the exact mathematical description of physical objects of rotation in system of the observer seem doubtful, because it is impossible to synchronize clocks between frames. It is clear also from the analysis of section \ref{sec:Struct}, that $ \psi (X, \tau) \neq \psi ' [X R (\tau)] $. We shall try to find the available methods of mathematical description.  

Let's consider a case of a periodic matrix of rotation $R(\tau)$ with the period $T=2\pi/\Omega$: $R(\tau) = R(\tau+2\pi/\Omega) $. As far as it is impossible to synchronize clocks between frames, we shall form some set of points $\Lambda$: $ \{ \tau_o+2\pi n/\Omega \}_{n=0}^{\infty}$ and will check the coincidence of functions $ \psi (X, \tau) \rightleftharpoons \psi ' (X ') $ on $\Lambda$. In these points according to expression (\ref{coord}) $X $ coincides with $X '$ (we shall write down it by definition as: $X ' = X_{|\Lambda}$). In a case examined by us it is possible to write also: $ \psi ' (X ') = \psi (X, \tau)_{| \Lambda}$\footnote{Generally, in case of symmetric function, this equality can be observed not only on set $ \Lambda $. Here for simplicity this opportunity is not considered.}. 

Any function of a kind: $ \psi (X, \tau) = \psi_o (X) \lambda (X, \Omega, \tau) + \mu (X, \Omega, \tau) $, where $ \psi ' (X ') = \psi_o (X)_{| \Lambda}, \, \lambda (X, \Omega, \tau)_{| \Lambda} =1 $ and $ \mu (X, \Omega, \tau)_{| \Lambda} =0 $ will satisfy to a condition of equality of functions corresponding to object of rotation in various systems of PhRD, since $ \psi ' (X ') = \psi (X, \tau) _ {| \Lambda} $. For such functions $ \lambda $ and $ \mu $ it is possible to present required function in the equivalent form:  

\begin{equation}
\psi (X, \tau) = \psi_o (X) \lambda (X, \Omega, \tau) + \psi_o (X) \mu (X, \Omega, \tau), 
\label {SRfunction}
\end{equation}

As far as the system of rotation should pass into system of the observer at frequency of rotation tending to zero: $ \psi (X, \tau) \rightarrow \psi (X ') $ at $ \Omega \rightarrow 0 $ functions $ \lambda $ and $ \mu $ should satisfy to additional conditions: $ \lim _ {\Omega \to 0} \lambda (X, \Omega, \tau) = 1 $ and $ \lim _ {\Omega \to 0} \mu (X, \Omega, \tau) = 0 $.  

The most simple way to satisfy to necessary conditions to functions $ \lambda $ and $ \mu $ is to demand: $ \lambda \equiv 1, \mu \equiv 0 $. It will correspond to a trivial case of non-rotating system of PhRD (rotating with zero frequency of rotation), coincides with the system of the observer. 

The elementary and non-trivial case to satisfy to necessary conditions, is to present $ \lambda $ and $ \mu $ as periodic functions of $ \tau $: $ \lambda (\tau) = \cos [\Omega_n (\tau-\tau_o)], \: \mu (\tau) = \pm i \sin [\Omega_n (\tau-\tau_o)] $ \footnote{Such approach exclude the spatial distribution and may not be suitable from being for all types of spatial rotations.}, where $ \Omega_n = n\Omega $, $n $ is an integer. We shall finally receive: 

\begin{equation}
\psi (X, \tau) = \psi_o (X) \exp [\pm i \Omega_n (\tau-\tau_o)]. 
\label{wavefunction}
\end{equation}

The given expression approximately describes an image, reflection of object of rotation in system of the observer. Remind, that we are examining rotations with motionless point of rotation, i.e. the image of object of rotation ``as a whole'' will be at rest in the system of the observer. Therefore: $ \tau-\tau_o =\int ^ {\tau} _ {\tau_o} d\tau = \int^b_a ds = - {\cal S}/\sigma $, where $ {\cal S} $ is action for some classical or quasi-classical physical object, $ \sigma $ is some constant (Landau and Lifschitz \cite {LL89}). In quantum mechanics such object is described with help of wave function: $ \Psi = {\cal A} \exp (i {\cal S}/\hbar) $, where $ \hbar $ is Planck's constant, $ {\cal A} $ is some ``slowly varying'' function. Correlation between this expression and expression (\ref {wavefunction}) is rather transparent. 

Comparing the function of object of rotation and the wave function of a free particle with rest mass $m_o$ it is possible to get $ \Omega=m_o c/\hbar $ and to come to a parity $ \hbar \omega = m_o c^2 $. Note, that expression (\ref {wavefunction}) will correspond to de Broglie's wave if to consider it from inertial $K^L$ to system of the observer $K$ frame (it is considered without loosing the generality $X^L=(x,y,z^L)$, $\gamma = 1/\sqrt{1-\beta^2}$, $\beta=V/c$): 

\begin{equation}
\fl \psi^L(X^L,\tau^L)=\psi_o(X^L)e^{i \left[ \Omega_n (\gamma-1)\tau^L-\Omega_n \beta z^L \right]} 
=\psi_o(X^L)e^{i \left( \Omega_n \gamma\tau^L-\Omega_n \beta z^L \right)}e^{i \left( \Omega_n \tau^L \right)}, \nonumber
\end{equation} 

if the term $e^{i \left( \Omega_n \tau^L \right)} $ is neglected. It means that the ``internal'' properties are neglected and are replaced with some ``macroparameters'' (energy, momentum, rest mass etc.). 

The description of the considered objects of spatial rotation (the set of all objects of rotation is much wider) correlates with the description of objects of quantum mechanics. From the mathematical point of view, this description is approximate because concurrences can be checked up only on countable set of points $ \Lambda $. Even at great values of $\Omega $ the information about function $ \psi'(X') $ can be exact enough, but is always incomplete. 

In points of set $ \Lambda $ function (\ref{wavefunction}) satisfies as group of rotations $ {\cal O}(3) $ and Lorentz group. According to our principle of compatibility entered in section \ref{sec:PhRD}, ``in a fashion standard of the quantum theory'' (L.Ryder \cite {Ryd85}; Landau, Lifschitz \cite {LL89}), entering the operators corresponding to macro-characteristics of object, it is possible to get the basic equations of quantum mechanics, such as Klein-Gordon, Schr\"{o}dinger and Dirac, as they are based on application of known space-time properties in system of the observer to the objects described by functions of a kind (\ref{wavefunction}). 

\section {Quantization of objects of rotation}
\label {sec:Quant}

Objects of rotation should visualize themselves as some ``influences'', known fields in system of the observer, otherwise it would not be possible to consider and try to describe them. Also, we want to find stable objects, and they cannot be an infinite energy source. There are two opportunities to satisfy to these conditions: to consider, that raised fields do not transfer energy or that fields are located in space. 

Let's assume, that the object of rotation described by function $\psi (X, \tau)$ (\ref {wavefunction}), in system of the observer $K$ can be a source of an electromagnetic field $u (X, \tau)$. Then this field should be described by the wave equation and corresponding Helmholtz equation (received with the substitution $u(X, \tau) = U (X) \exp (i\Omega\tau)$): 
\begin{equation}
\Box u(X, \tau) = \psi (X, \tau), \: \nabla^2 U (X) + \Omega^2 U (X) = - \psi_o (X). 	
\label{waveequation}
\end{equation}

There are partial solutions of Helmholtz equation located in space. For example, in one-dimensional case for $\psi_o (x) = \delta (x+R) \pm \delta (x-R)$ there are solutions (an even mode): $U^{in}_{even} (x) = \pm \exp (\pm i\Omega R) / (i\Omega) \cos (\Omega x) $ for ``inside'' area $ |x | <R $ and $U^{ex}_{even}(x) = \pm \exp (\mp i\Omega x) / (i\Omega) \cos (\Omega R)$ for ``external'' area $|x |> R$. At $ \Omega R =-\pi/2 + \pi n $: $U^{out}(x) = 0$ for $|x |> R$, while $U^{in} (x) \neq 0$ for $|x | <R$. 

Similar solutions of the Helmholtz equation (\ref{waveequation}) can be found in $2D-$ and $3D-$ case. And solutions exist in classes of both continuous and discontinues functions. For example in 3D-case for $\psi_o (r) = \delta (r-R) / (4 \pi R^2)$ solutions of Helmholtz equation from a class of continuous functions are: $U^{in}_{even}(r) = - i e^{-i\Omega R} \cos (\Omega r) / (\Omega R r), \; U^{ex}_{even}(r) = - i e^{-i \Omega r} \cos (\Omega R) / (\Omega R r)$ for which fields appear located in space under condition of $\Omega R = -\pi/2 + \pi n$. Such conditions we shall name conditions of quantization. 

\begin{figure} [h] 	
\begin{center}
		 \includegraphics*[angle=270, width=60mm] {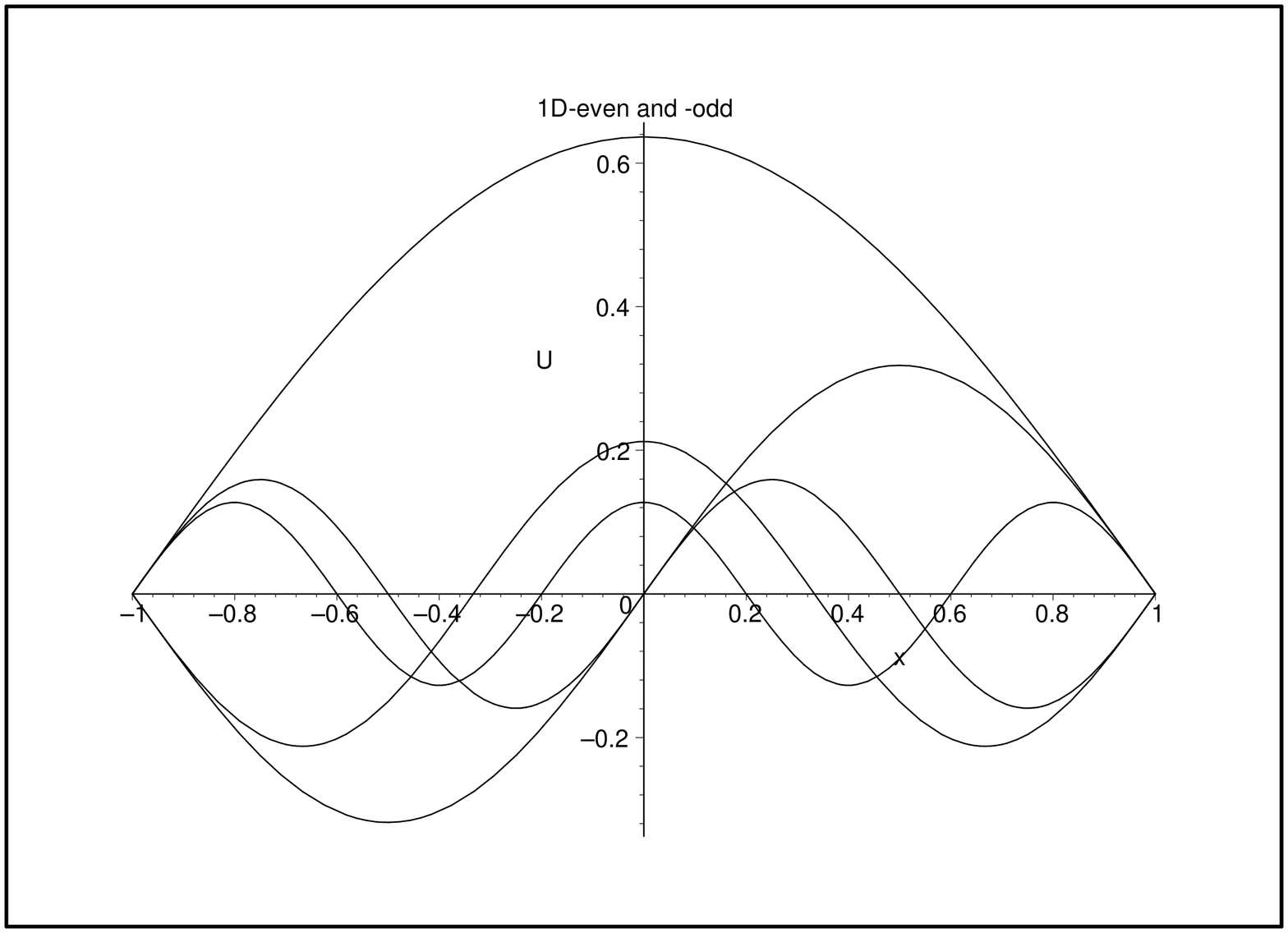}
		 \includegraphics*[angle=270, width=60mm] {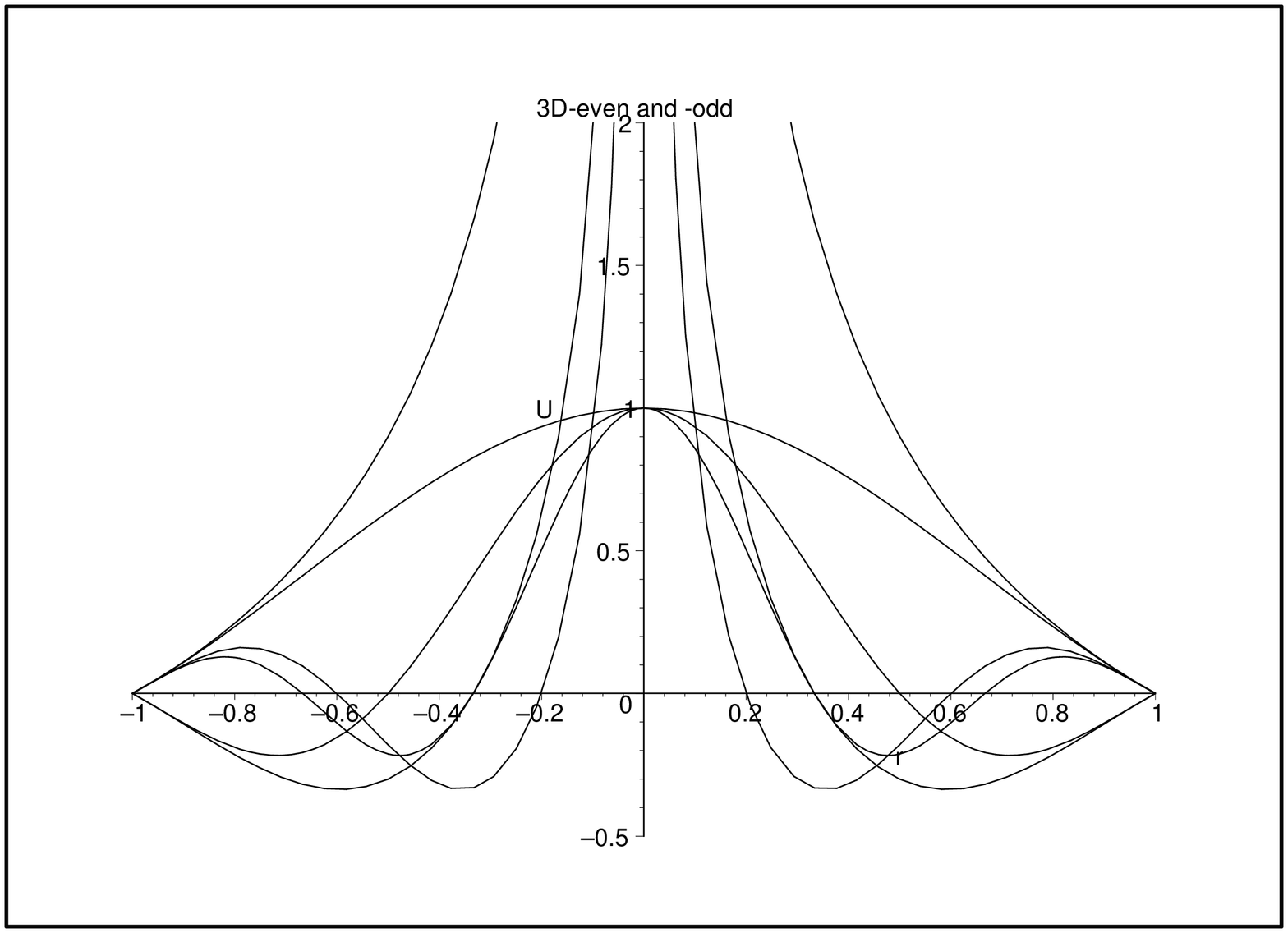}
	 \caption{Amplitudes of the excited fields of object of rotation ($1D$ and $3D$).}
	 \label{fig:ObjDistr}
	 \end{center}
\end{figure}

Thus, the energy conservation law in system of the observer can be satisfied for the fields excited by object of rotation. Necessary conditions of this are conditions of quantization of frequencies and spatial sizes of objects of rotation. The distributions of amplitudes of the excited fields $U (X)$ (even and odd modes) from a class of continuous functions for $1D-$ and $3D-$ spaces are presented on Fig. \ref{fig:ObjDistr}. 

Analysis shows that in three-dimensional case it is possible to find higher order solutions of the Helmholtz equation (\ref{waveequation}) from the class of continuous functions: 
\begin{equation}
\fl U(r,\theta,\phi) =  \sum_{l=0}^{\infty}\sum_{m=-l}^{l} R_{lm}(r)Y^{m}_{l}(\theta,\phi), \,
R(r)  =  \frac {C_{1}}{\sqrt {r}}J^{(1)}_{l+\frac {1}{2}}(\Omega r)+ \frac {C_{2}}{\sqrt {r}}J^{(2)}_{l+\frac {1}{2}}(\Omega r), \label{Sol}
\end{equation} 

where $C_{i}$ are some constants, $J^{(i)}_{l+\frac {1}{2}}(\Omega r)$ are Bessel's functions of the first and the second kind correspondingly. As far as, the function $U(X)$ needs to be zero in outside region, at $\|X\|>R$, so one can get the necessary condition for the stable object of rotation in $K$: $J^{(1)}_{l+\frac {1}{2}}(\Omega R)=\, 0$, for even mode and $J^{(2)}_{l+\frac {1}{2}}(\Omega R)=\, 0$ for odd mode. So, the objects of rotation in $K$ are allowed to have discrete `` sizes'' $R_{i}=\, \alpha_{i}/\Omega $, where $\alpha_{i}$ are the zeros of the Bessel functions. 

\begin{figure} [h]
	 \begin{center}
	 \includegraphics*[angle=270, width=60mm] {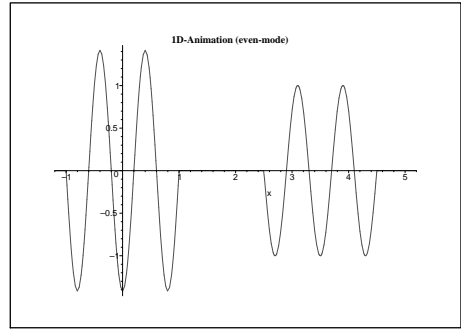}
	 \caption{Formation of object of rotation.}
	 \label{fig:ObjEst}
	 \end{center}
\end{figure}

Helmholtz equation gives the steady-state solutions. The analysis of solutions of the wave equation (\ref {waveequation}) shows, that at the beginning, during the process of ``formation'' of object of rotation (or changing of its state) some objects ``departing from it'' with speed of the excited waves are formed also. This process is shown on Fig. \ref{fig:ObjEst} in 1D- case. 

Thus, the stable object of rotation can not radiate energy in system of PhRD of the observer. Necessary conditions for this are that its spatial sizes and frequencies of rotation are quantized. The stable state of object of rotation corresponds to the fixed frequency of rotation, i.e. the uniformly rotating system. Transition between stable states of object of rotation is accompanied by formation of "departing" objects in system of the observer. 

\section{Objects of spatial rotation in micro- and macro-scales} 
\label{sec:MMScale}

It was mentioned above in Sect.\ref{sec:PhRD} that the principle of relativity is valid in general theory of relativity both for any local region (in micro-scale) and far from the gravitating bodies (in macro-scale). It is one of the basic GTR principles. Exactly in these regions our ``new'' physical objects may find quite interesting physical interpretation. 

The description of objects of rotation in micro-scale (in the field of elementary particles and physical vacuum) is the description of their properties from the point of view of the external observer, in relation to system of rotation. 

From this point of view the properties of non-observable objects of rotation (in external regions) in many respects coincide with objects of quantum mechanics. Functions such as wave functions correspond to objects of rotation in the system of the observer, and they satisfy to the equations of quantum mechanics (Klein-Gordon, Schr\"{o}dinger and Dirac). States of objects of rotation are quantized, some of them are inherent in spin characteristics. 

Observable objects of rotation (in internal regions) can be identified in system of the observer, hence, can interact with other physical objects. Observable objects seem ``captured'' inside the area bounded by horizon of events. From system of the observer it can be looked as the phenomenon of confinement. In quantum field models, the new type of interactions - the strong one was introduced in addition to gravitational and electromagnetic interactions to explain this phenomenon. It was not a consequence of the physical worldview at that time and, of course, this new interaction needed to be declared as a fundamental one. 

The suggested model could be checked up on the experiment proposed by Academician of RAS G.T.Zatsepin in interaction of massive particles or nucleus of atoms with high energy massive particles. If time of interaction $t_c$ is less than the minimal period of internal rotation of a target particle $T=2\pi/\omega=2\pi\hbar / (M_o c^2) $, than particles produced during interaction should lie in the same spatial plane. Energy of interaction may be estimated as: ${\cal E} = m_o c^2/\sqrt{1-\left (2 \hbar / M_o c^2 t_c\right) ^2}$. Anticipatory estimations show, that energy of particles should be great enough, not less than tens GeV\footnote{This effect has been already observed few years ago by G.T.Zatsepin group by product during investigation of cosmic particles interactions, but these results were not published, because had been seemed unusual and had not had satisfactory explanations}. 

Interesting correlations with topological ideas declared in this paper may be found in works of Y.Lin \cite{Liu00} in his investigations of the weak CP phase. 

Quantized electromagnetic fields excited by objects of rotation (see section \ref {sec:Quant}) are good candidates for a role of electroweak interactions. Their general nature is already proved experimentally. It seems that both weak and strong ``fundamental'' interactions are straight consequences of ideas declared above. 

In this paper, the physical objects were mainly investigated in micro-scale. Very interesting development of declared ideas is being looked through in macro-scale. The system of the observer may also be considered as system of rotation. This way the observer is located inside the horizon of events of system of rotation. Note, that the horizon of events exists only for observer from different system of PhRD. The horizon of events can not be perceived by observer in his own system of PhRD. The non-observable objects will be perceived by observer as a strange substance changing our visible space-time topology. Now, good candidates for such substance exists in astronomy. It is so-called dark matter, dark energy. The quantized fields excited by these objects may be perceived as streams of particles penetrating our Universe in different directions. Such sight at objects of rotations in macro-scale may appear useful at the consideration of nowaday problems of astrophysics.  

\section*{Acknowledgements}
I would like to express the sincere gratitude to Academician of the Russian Academy of Science G.T.Zatsepin for moral support of this work and gratitude to Professor L.H.Ryder for useful discussions. 

\section*{References}


\begin{thebibliography}{99}

\bibitem{Poin04} Poincar\'e H 1904 {\it Bull.de Sciences Math} {\bf vol.28, ser.2}, p.302; 1905 {\it The Monist} {\bf vol. XV}, p.1.  

\bibitem{Log89} Logunov A~A, Mestvirishvilli M~A 1989 Relativistic Theory of Gravitation  {\it Nauka, Moscow}.  

\bibitem{LL88} Landau L~D, Lifschitz L~M 1988 Theoretical Physics: The Classical Theory of Fields {\bf vol.2} {\it Nauka, Moscow}. 

\bibitem{Ryd85} Ryder L~H 1985 Quantum Field Theory {\it Cambridge Univ. Press}. 

\bibitem{LL89} Landau L~D, Lifschitz L~M 1989 Theoretical Physics: Quantum mechanics. Non-relativistic theory {\bf vol.3} {\it Nauka, Moscow}. 

\bibitem{Liu00} Liu Н 2000 Geometric origin of the weak CP phase {\it Phys. Review {\bf D}}, Vol.{\bf 61}, 033010.

\end {thebibliography}
\end {document}